# AN EFFICIENT GROUP KEY MANAGEMENT USING CODE FOR KEY CALCULATION FOR SIMULTANEOUS JOIN/LEAVE: CKCS

Melisa Hajyvahabzadeh[1], Elina Eidkhani[1],
S. Anahita Mortazavi[1], and Alireza Nemaney Pour[2]

[1]Dept. of IT Engineering, Sharif University of Technology, Kish Island, Iran
`melisa.vahabzadeh@gmail.com; elina.eidkhani@gmail.com;`
`mortazavi.anahita@gmail.com`
[2]Dept. of Computer Software Technology Engineering,
Islamic Azad University of Abhar, Iran
`pour@abhariau.ac.ir`

## ABSTRACT

*This paper presents an efficient group key management protocol, CKCS (Code for Key Calculation in Simultaneous join/leave) for simultaneous join/leave in secure multicast. This protocol is based on logical key hierarchy. In this protocol, when new members join the group simultaneously, server sends only the group key for those new members. Then, current members and new members calculate the necessary keys by node codes and one-way hash function. A node code is a random number which is assigned to each key to help users calculate the necessary keys. Again, at leave, the server just sends the new group key to remaining members. The results show that CKCS reduces computational and communication overhead, and also message size in simultaneous join/leave.*

## 1. INTRODUCTION

IP multicast (hereinafter multicast) is an efficient group communication protocol to deliver multicast content from a single source to multiple users. This communication technology uses IGMP (Internet Group Management Protocol) [1] for group membership that allows members to join the group and receive content freely. As a result, open group membership by IGMP leads to eavesdropping. In order to avoid this threat, group key management has been proposed. Group key is a key that is shared by all group members and the sender for encrypting data by the sender and decrypting transmitted data by the group members.

The security requirements of secure multicast are forward secrecy and backward secrecy [2]. Forward secrecy ensures when a member leaves the group, he/she cannot access successfully the current content. Backward secrecy ensures that a new member cannot access any data which is sent before its join process. Because of these requirements, group key needs to be updated on each membership change and to be distributed to the valid group members securely. This process is called group re-keying or re-keying in short.

Two types of group membership, single and simultaneous exist in multicast. Figure 1 illustrates the process of join/leave in which only one member, $u^{n+1}$ joins/leaves the multicast group. In single join/leave, the server is considered to reply only one user's join/leave request. However in real world, simultaneous join/leave occurs more frequently. Simultaneous join/leave refers to multiple requests from different users to the server at the same time. Figure 2 shows the simultaneous join/leave operations which occur concurrently in a multicast group. In simultaneous join/leave, the server needs to response multiple requests synchronously. As providing forward/backward secrecy is important in single join/leave, these requirements are crucial in simultaneous case as well.

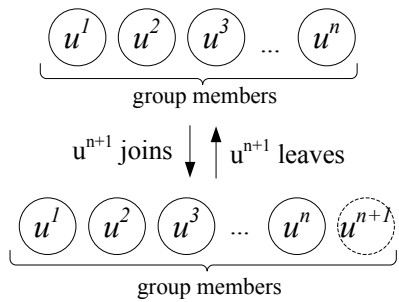

Figure 1.  Single join/leave

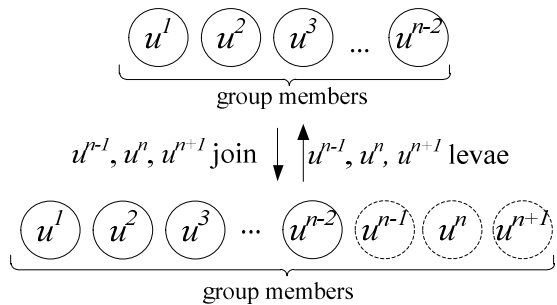

Figure 2.  Simultaneous join/leave

Re-keying implies large overhead on each membership change for dynamic groups. Usually, the overhead at leave is larger than the overhead at join. Because when a new member joins the group, the new group key can be encrypted by the previous one and distributed by multicast to the existing members except the new member who receives the group key by unicast being encrypted by his/her individual key. But when a member leaves the group, the group key is encrypted by each member's individual key and is transmitted by unicast to the remaining members individually. In this case, the previous group key is supposed compromised. So, for a group with $n$ members, the re-keying overhead is $O(n)$.

Many protocols have been proposed for group key management [3]-[10]. The main purpose of them is how to distribute group key to valid members efficiently at leave. The proposed protocols are based on logical hierarchy model. The common issue with all of these protocols is that they focus only on single join/leave. Although these protocols reduce the re-keying overhead largely at leave from $O(n)$ to $O(log\ n)$, they increase re-keying overhead from $O(1)$ to $O(log\ n)$ at join [3]-[8]. While proposals in [9],[10] reduce re-keying overhead at join for multicast communication compared with [3]-[8], none of them [3]-[10] focus on reducing re-keying overhead for a new member at join.

In fact, reducing re-keying overhead for new members at join is an important factor for simultaneous mode. Because, when $m$ members join the group simultaneously, the server should deliver the necessary keys to both the new and the current members. For the current members, the overhead has already been decreased, but for new member the overhead still remains. So, in simultaneous mode, reducing overhead of key delivery to new members at join is a critical factor. Therefore, it is necessary to have the minimum overhead for the new members at join.

For this purpose, the overhead of re-keying for new member in single join/leave should be minimized at first. Then, by improving that method, the proposed protocol can be used in simultaneous mode.

As stated before, most of the previous approaches do not consider simultaneous join/leave. [11] is the only research that addresses simultaneous join/leave by periodic re-keying. However, this proposal does not support forward and backward secrecies. This is the main problem with this protocol because some members may repeat join/leave within a period. Moreover, this approach does not propose any specific method for re-keying.

This paper develops the mechanism of key tree management in [12] and adopts the protocol to simultaneous join/leave. CKCS is a hierarchical protocol which has the minimum re-keying overhead at join/leave compared to [3]-[10]. In CKCS, all necessary keys are calculated by members using code for key calculation rather than distributed by the key server. As a result, key generation, key encryption and message size overhead are reduced to $O(m)$ where $m$ denotes the number of members who join the group simultaneously.

This paper is organized as follows. Section 2 shows the overview of the secure multicast network, and discusses the researches which are based on the hierarchical approach. The design principles and detailed design of our proposal are shown in sections 3 and 4, respectively. At the end of section 5, we discuss the security of our protocol. In section 6, we compare our protocol with the hierarchical based researches. Section 7 is the conclusion.

## 2. RELATED WORK

### 2.1. Network Structure

Figure 3 illustrates the network structure of secure multicast. This network has three major parts; key server, multicast sender and multicast members. Key server is responsible for generating the keys and delivering them to both multicast sender and the existing members. Multicast sender encrypts the contents by the key which is received from the key server and distributes the encrypted content to all the group members. Multicast members are the authorized users who receive group content from multicast sender and also the necessary keys from the key server.

When a new member joins a multicast group, he/she should send IGMP request to his/her nearest router to receive multicast data from the network. Also, the new member needs to send a join request to the key server. When the new member's request is accepted by the key server, re-keying process should be started. All the updated keys must be delivered by the key server to multicast sender, the new member and all the existing members. When a member leaves a multicast group, he/she needs to send IGMP leave message to stop content delivery at first. Then, this member should inform the key server by sending a leave message. In next subsection, re-keying process of some previously proposed protocols is reviewed.

### 2.2. Existing LKH Approaches

Wallner et al. in [3] proposed LKH (Logical Key Hierarchy) approach for group key management in secure multicast. Later, several versions of LKH based protocols were proposed [4]-[10]. In these protocols, a centralized server is responsible for managing the group and distributing the group key to all group members. In LKH based protocols, the members of multicast group are mapped to the leaves of a logical key tree. This tree is a d-ary tree, typically a binary one. Each member stores all the keys from the leaf node along the path to the root. The root key is the group key. When a member joins or leaves the group, all the keys in his/her possession need to be changed to new ones. The lowest level in the tree can be the starting point in a way that a new parent key is encrypted by using its two child keys and distributed to users. Since the height of the key tree is *log n*, as a resultthe complexity of re-keying is also $O(log\ n)$.

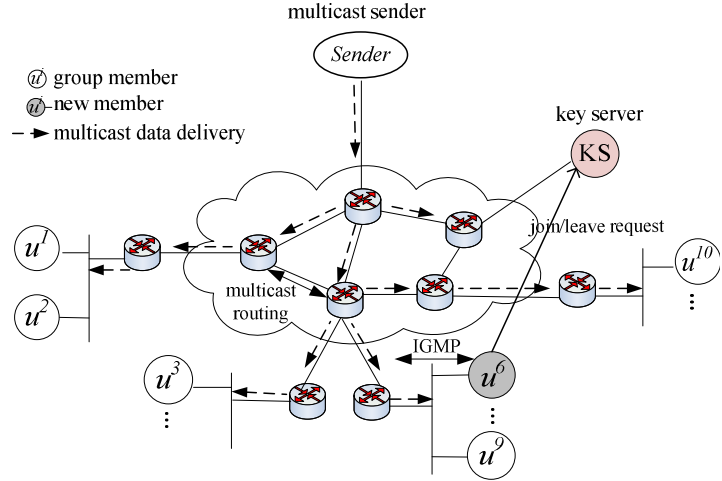

Figure 3. Network structure of secure multicast [9]

Figure 4 shows the outline of re-keying procedure for LKH at join. This figure depicts the logical key tree with seven existing members, $u^1$ through $u^7$, when $u^8$ joins the group. At this time, the affected middle node keys, $K_{7,8}$and $K_{5,8}$, and the group key, $K_G$, are changed to $K'_{7,8}$, $K'_{5,8}$, and $K'_G$ respectively, because these nodes are on the path from $u^8$ to the root. Then, the key server sends the new keys to the related members. First, all the keys which $u^8$ needs to have, $K'_{7,8}$ , $K'_{5,8}$and $K'_G$ are sent by unicast to $u^8$ being encrypted with $K_8$, $K'_{7,8}$and $K'_{5,8}$, respectively. Next, $K'_{7,8}$is sent to $u^7$ being encrypted by $K_7$. For the existing members, the updated keys are sent by multicast. In this example, $K'_{5,8}$is sent to $\{u^5,u^6\}$, and $u^7$ being encrypted with $K_{5,6}$ and $K'_{7,8}$respectively. $K'_G$ is sent for $\{u^1,u^2, u^3, u^4\}$ and for $\{u^5, u^6, u^7\}$ being encrypted with $K_{1,4}$ and $K'_{5,8}$, respectively.

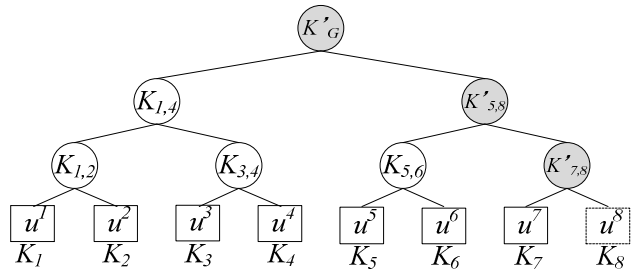

Figure 4. An example of logical key tree for LKH and OFT at join/leave

For leave operation, the same procedure is performed. When a member leaves the group, for example $u^8$, the node key of the leaving member should be deleted from the key tree at first. Next, all the affected middle nodes in the path to the root are updated to new ones and are sent to remaining members. By this method, LKH reduces re-keying overhead at leave from $O(n)$ to $O(log\ n)$.

OFT [4] is another LKH based approach. The main idea of this approach is to reduce the key distribution overhead of the server by shifting a part of key calculation to users' side. The key of node ($i$), $K_i$, is calculated by using the formula $K_i = f(g(K_{left\,(i)}), g(K_{right\,(i)}))$. In this formula, $f(x)$ is a mixing function and $g(x)$ is a one-way hash function. The value of $g(x)$ is called blinded key. $K_{left\,(i)}$ denotes the key of left child while $K_{right\,(i)}$ denotes the key of right childe of the node.

In OFT, when a member joins a multicast group, he/she receives some information includes the group key, his/her sibling's blinded key, and ancestors' sibling blinded keys. For example, in Figure 4 when $u^8$ joins the group, he/she should receive the sibling blinded key, $g(K_7)$, and his ancestors' siblings blinded keys which are $g(K_{5,6})$ and $g(K_{1,4})$. These keys are encrypted by $K_8$. On the other hand, $u^8$ calculates $K'_{7,8}$, $K'_{5,8}$ and $K'_G$ by using the following formulas.

$$K'_{7,8} = f\left(g\left(K_7\right), g\left(K_8\right)\right),$$
$$K'_{5,8} = f\left(g\left(K_{5,6}\right), g\left(K_{7,8}\right)\right), \quad\quad\quad (1)$$
$$K'_G = f\left(g\left(K_{1,4}\right), g\left(K'_{5,8}\right)\right).$$

After $u^8$ joins the group, the blinded keys of $K'_{7,8}$, $K'_{5,8}$ and $K'_G$ are encrypted with their sibling keys and advertised by multicast to the existing group members as follows:

$$s \xrightarrow{multicast} \begin{cases} u^1 - u^4 : \left(g\left(K'_{5,8}\right)\right)_{K_{1,4}} \\ u^5 - u^6 : \left(g\left(K'_{7,8}\right)\right)_{K_{5,6}} \\ u^7 : \left(g\left(K_8\right)\right)_{K_7} \end{cases} \quad\quad (2)$$

Now, by receiving these blinded keys, all the group members can update the necessary keys through the above formula. In this method, when a member leaves the group, the key updating process will be done as well as join operation. Therefore, OFT reduces the overhead of re-keying from $O(log\,n)$ to $O(1/2\,log\,n)$.

In [9], authors divide the group members into some subgroups and assign a subgroup key to each subgroup. In this protocol, the key server assigns a secret to each joining member, and the group key is generated using those secrets. Moreover, each member is given the inverse value of secrets of all the other members except its own. Those inverse values help the remaining members to update the group key at leave. When a member leaves the group, the key server only informs the remaining members that the member has left the group. The new key is generated by individual members using the inverse value of the leaving member. In order to reduce the burden of maintaining the inverse values, this protocol divides the current members into subgroups. By this mechanism, the protocol can reduce the computational and communicational overhead at member leave and the maintenance overhead of inverse values at individual members. Although this protocol focuses on reducing the re-keying overhead at leave but it also reduces re-keying overhead at join. Consequently, this protocol reduces re-keying overhead at leave from $O(log\,n)$ in LKH to $O(log\,[n/m])$ at leave when $m$ is the number of subgroups.

OKD [10] is defined based on one-way key derivation. In this method, users compute some of the updated keys in each group membership change instead of receiving all new keys from the server. A derivation function, $f(x)$ is used to generate new keys from the old ones. OKD considers 3-ary tree for key tree. In OKD, when a new member joins the group, he/she is assigned to a suitable branch of the key tree and all the necessary keys are sent by unicast to the new member but the current group members calculate their necessary keys by themselves.

While OKD reduces re-keying overhead at join for multicast communication compared with other LKH based protocols [2]-[7], it does not focus on reducing re-keying overhead for a new member at join. So, OKD is not suitable when several members join a group simultaneously. The details of OKD exist in [10].

As shown above, LKH based methods are acceptable because they reduce re-keying overhead largely. If the group size is small, for example less than hundred members, one might not use a hierarchical approach. However, when the group size grows to several thousands or millions, hierarchical approaches such as LKH based protocols are needed. More important, when simultaneous join is considered for a system, the overhead of re-keying for the new members is crucial because the server has to generate the new keys and encrypt them for each new member individually to send them. While protocols in [9],[10] reduce re-keying overhead at join for multicast communication compared to the other LKH based ones [3]-[8], they do not focus on reducing re-keying overhead for the new members which is necessary for simultaneous join. Therefore, the overhead of re-keying for new members in simultaneous mode is a crucial factor. To handle this issue, we develop the mechanism of the key tree for simultaneous join/leave in our previously proposed protocol [12] and now we evaluate its performance in simultaneous mode.

## 3. DESIGN PRINCIPLE

We now present our new efficient protocol for group key management in simultaneous mode. In fact, this work (CKCS: Code for Key Calculation in Simultaneous), is an extension of our original idea [12]. We add some features and adopt them to simultaneous mode. The following design principles have been performed in our protocol.

(1) Figure 3 illustrates the network structure of secure multicast. Using the same network structure for CKCS, the key server is responsible for generating and distributing only the group key to new members at join. First of all, new users send an IGMP message to the nearest router for receiving the multicast content from a multicast sender. Next, new users need to send a join request to the key server for receiving the session group key for decrypting the encrypted content. Finally, when members leave the group, they send the leave requests to the key server. The key server updates the group key and redistributes it to the remaining members and multicast sender by multicast.

(2) Figure 5 illustrates the logical key tree in CKCS. The tree applies the concept of key tree in LKH approaches. When $m$ members join a group simultaneously, the server creates a key tree for those users, and combines it with the current key tree by adding a new node to top of these two trees. Here, this top node is considered as the group key. By this technique, the height of the key tree remains unchanged and as a result we have a balanced tree.

(3) Similar to LKH based approaches, in CKCS, some keys need to be updated after each membership change. The updated keys are calculated by the group members rather than distributed by the key server. For this purpose, each node of the key tree is assigned to a specific code called node code. A node code is a random number which is assigned to each middle node key to help the users calculate the necessary keys. This code is delivered to the new member at join as a position of that member in the key tree. It is calculated by concatenating a random number to right digit of its parent node code. Generally, each parent node code can be obtained by deleting the rightmost digit from his/her child code. By this mechanism, each member knows the codes of all nodes in his/her path to the root. So, members can update the affected node keys using these codes and the hash function after each membership change (Figure 6).

(4) In this step, we explain simultaneous join in CKCS. When several users join a group concurrently, the server creates a new key tree for new users by assigning a position code to the top node of the new key tree. The previous key tree and the new one are concatenated to each other by adding a top node which is considered as a new group key. The key server encrypts the new group key with each new user's individual key and sends them by one multicast message. Each new member generates the middle node keys in his/her path to the root by applying hash function on bitwise XOR of the group key with each node code. Current members only need to compute the top node of new key tree which is calculated by applying hash function on the previous group key.

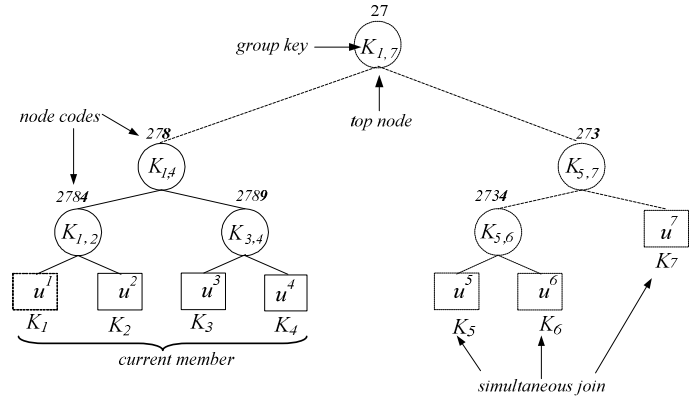

Figure 5.  The logical key tree in CKCS

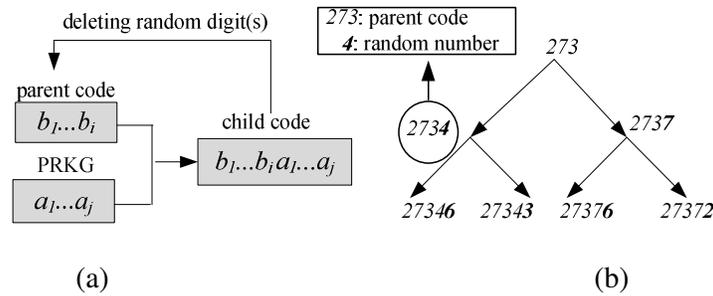

Figure 6.  Node code management (a) node code generation (b) simple example

(5) This step describes simultaneous leave in CKCS. Depending on the position of leaving members, two cases are considered; the worst case and the best case (Figure 7). The worst case occurs when members leave the group from different leaves of the key tree while the best case happens when members who are located in one half of the key tree. At leave, the sibling node of each leaving member is moved to his/her parent position in the key tree. To provide forward secrecy, the server is responsible for updating the group key and sending it to the remaining members. The server encrypts the new group key with top node of each part and sends it by multicast to all remaining group members. The re-keying overhead at leave has the lowest value by considering the best case.

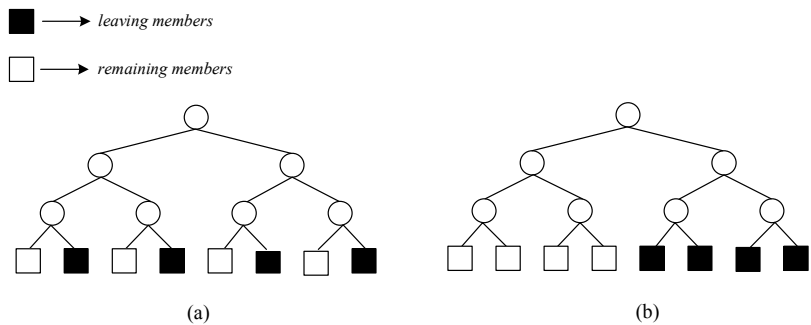

Figure 7.  Leave operation in simultaneous mode (a) worst case (b) best case

## 4. DETAILED DESIGN

In this section, we present CKCS in details. CKCS focuses on user side key calculation rather than server side key distribution. In this protocol, a code is assigned to each middle node of the key tree to help users calculate the necessary middle node keys.

### 4.1. Key Tree Structure and Node Code Management in CKCS

The key server manages the group members and sends them only the group key. When new members join the group, they need to know their individual keys and their positions in the key tree according to the basic information. After receiving join requests, the key server sends the new members their individual keys, and also their position codes in the key tree. The individual keys of the new members and their position codes are sent via a secure channel. Each node of the key tree has a unique code. New members can calculate the necessary middle node keys in their path to the root by using node codes. Each middle node key is updated by applying hash function on bitwise XOR of the group key and the corresponding node code as below.

$$K_{middle\_node} = f(K_G \oplus node\_code) \qquad (3)$$

Moreover, the group key is updated after each membership change. New members receive the group key from the key server encrypted by their individual keys. The current members can calculate it by applying one-way hash function on the previous group key as below.

$$K'_G = f(K_G) \qquad (4)$$

In simultaneous join, when multiple users join the multicast group, the server creates the key tree of new users at first and allocates each new user to one leaf of the key tree. The key tree of old group members and new simultaneous users are combined with each other by adding a new node to the top. At this time, the top node is assumed as the new group key. The code of the new top node (root node) is calculated by deleting a digit from the rightmost digit of the previous root code. Then, the codes of middle nodes for simultaneous join are created by concatenating a random number to the rightmost digit of the parent code as below.

$$Childnode\_code = (Parentnode\_code \parallel Random\ digit(s)). \qquad (5)$$

Figure 8 shows the node code management procedure in simultaneous mode. Here, by joining the new users, $\{u^5, u^6, u^7, u^8\}$, simultaneously, new node key, $K_{1,8}$, is created on top of the previous root node key, $K_{1,4}$. The key tree of new members, $\{u^5, u^6, u^7, u^8\}$, is located to the right child position of the new root node. Here, the code of new root is $27$ calculated by deleting $8$ from the right side of $278$. The middle node codes of doted branches (key tree of new members) are created by attaching a random number to the right side of their parents' code.

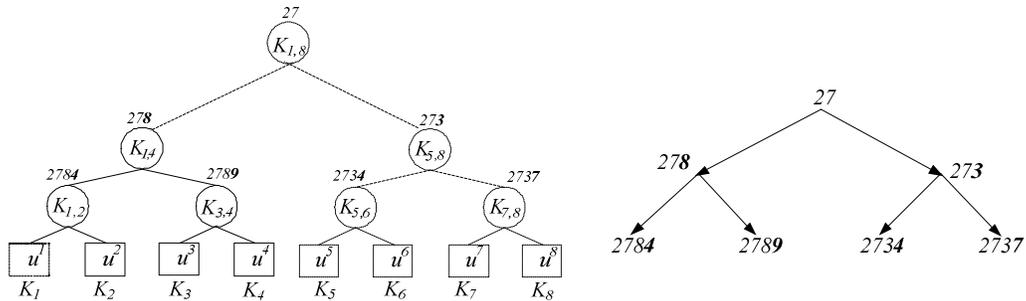

Figure 8.  Node code management key tree in simultaneous case

## 4.2. Join Operation

We use Figure 9 to explain how re-keying is done in simultaneous join by a simple example of a multicast group with *4* current members $\{u^1, u^2, u^3, u^4\}$ while $\{u^5, u^6, u^7\}$ join the group simultaneously. The procedure is done as follows:

(1) $\{u^5, u^6, u^7\}$ send join requests to the key server concurrently. The server creates their key tree and generates their individual keys. Individual keys of the new members, $\{K_5, K_6, K_7\}$, are sent through a secure channel to them. The key tree of new members is attached to the current key tree by adding a top node, $K_{1,7}$, to the top of $K_{1,4}$. This new top node, $K_{1,7}$, is the new group key.

(2) There server calculates code of $K_{1,7}$ and all the new middle nodes in the new key tree. *27* is assigned to $K_{1,7}$ by deleting *8* from the rightmost digit of *278* assigned to $K_{1,4}$. In the new members' key tree, code of each middle node is generated by attaching a random number to the rightmost digit of his/her parent node code. These codes are *273, 2734* assigned to $K_{5,7}$ and $K_{5,6}$, respectively. After generating these nodes codes, the server sends the position codes to them through a secure channel.

(3) The key server updates the group key from $K_G$ to $K'_G$, using one-way hash function.

$$K'_G = f(K_{1,4}) \qquad (6)$$

(4) Then, $K'_G$ is encrypted by each new member's individual key and sent by one multicast message.

$$s \xrightarrow{\text{multicast}} \{u^5, u^6, u^7\} : (K'_G)_{K_5}, (K'_G)_{K_6}, (K'_G)_{K_7}. \qquad (7)$$

(5) The current group members, $\{u^1, u^2, u^3, u^4\}$, who are located in current key tree, calculate the new group key, $K'_G$, by applying one-way hash function to the previous group key $K_G$.

$$u^1, u^2, u^3, u^4 : K'_G = f(K_{1,4}) \qquad (8)$$

(6) The new members, $\{u^5, u^6, u^7\}$, compute all their necessary middle node keys in their path to the root by the following formula:

$$u^5, u^6 : K_{5,6} = f(K'_G \oplus 2734),$$
$$u^5, u^6, u^7 : K_{5,7} = f(K'_G \oplus 273). \qquad (9)$$

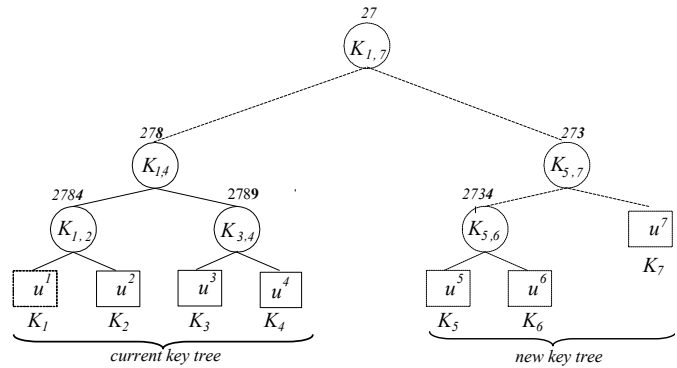

Figure 9.  Updating key tree in simultaneous join in CKCS

### 4.3. Leave Operation

In CKCS, when several members leave the group, the key tree is divided into two equal parts. As Figure 10 shows, the deleted nodes belong to one half of the key tree. So, these parts will be divided again and again until just the leaving members' branches are remained. The number of tree divisions is equal to *(log n -1)* where *n* is the number of group members. The key server encrypts the new group key by the keys of the top node of each half. We now use Figure 10 to explain how re-keying is done in simultaneous leave by a simple example of a multicast group with *8* members *{u¹, ... , u⁸}* when *{u¹, u⁴, u⁸}*leave the group. The procedure is done as follows:

(1) When *{u¹, u⁴, u⁸}* leave the group, the nodes of $K_{1,2}$, $K_{3,4}$, and $K_{7,8}$ are deleted from the key tree and $u^2$, $u^3$ and $u^7$ are promoted to their top node positions.

(2) The key server generates a random group key, $K'_G$.

(3) $K'_G$ is sent to the remaining group members by multicast, being encrypted by the top node key of each part, $K_2$, $K_3$, $K_{5,6}$, $K_7$. The group key is sent to the members who are located in each part, part-1, part-2, part-3, and part-4 respectively.

$$s \xrightarrow{multicast} \begin{cases} u^2 : (K'_G)_{K_2} \\ u^3 : (K'_G)_{K_3} \\ u^5, u^6 : (K'_G)_{K_{5,6}} \\ u^7 : (K'_G)_{K_7} \end{cases} \tag{10}$$

(4) Now, *{u², u³, u⁵, u⁶, u⁷}* whose their middle nodes are affected from this simultaneous leave, update $K_{1,4}$ and $K_{5,7}$to $K'_{1,4}$and $K'_{5,7}$respectively by applying one-way hash function on bitwise XOR of the group key and the related node codes.

$$u^2, u^3 : K'_{1,4} = f(K'_G \oplus 278),$$
$$u^5, u^6, u^7 : K'_{5,7} = f(K'_G \oplus 273). \tag{11}$$

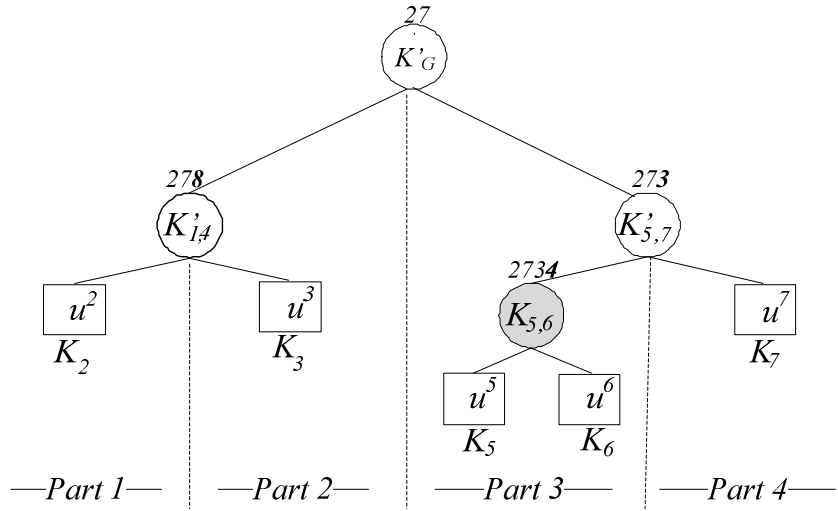

Figure 10. Simultaneous leave for CKCS

## 5. SECURITY ANALYSIS

In this section, we analyze the security requirements of CKCS. Backward secrecy ensures that new members at join cannot achieve the archived contents in the group. In CKCS, the new group key is organized by applying one-way hash function to the previous group key. One-way hash function has the property of one-wayness which means that it is easy to calculate $y = E(x)$, but by given y it is computationally hard to find $x$. Therefore, it is impossible for a new member to find the previous group key.

Forward secrecy ensures that when several members leave the group, they cannot access successfully the current contents. In CKCS, the new members cannot generate the current session keys with their previous information because the group key is generated by the key server at leave, and is transmitted by the node keys that the leaving members do not have them.

## 6. COMPARISON

In this section, we compare CKCS protocol with some previously proposed ones, LKH, OFT, and OKD. We compare these protocols at join and leave operations for simultaneous mode. The comparison measures are based on key generation, key encryption, communication overhead, and message size. Tables 1, 2, 3 and 4 summarize our comparisons, focusing on the following measures:

- Computational overhead
  - Key generation overhead: the number of keys that must be generated at join/leave.
  - Encryption overhead: the number of encryptions.
- Communication overhead: the number of transmissions from the key server.
- Message size: the total number of keys in one message.

The computational overhead is the sum of key generation and key encryption. In our comparisons shown in the following tables, $n$ denotes the group size which is the number of members in the multicast group after join and before leave operations. In addition, in simultaneous mode, $m$ denotes the number of members that join or leave multicast group concurrently. Finally, in simultaneous mode we consider that $m \leq n$. In other words, the number of simultaneous users is less than or equal to the number of group members.

Binary key tree is assumed for key degree in comparison. As stated before, in binary key tree the height of tree is $log_2 n$ which shows the number of nodes in each branch. Obviously, the efficiency of a protocol is related to the height of the key tree. In other words, a key tree with smaller height is more efficient than a tree with larger height. Consequently, since the tree height for all of these protocols is equal, the factors that make differences in decreasing overhead are the re-keying procedure and the key distribution technique.

The re-keying method itself is an effective way to reduce the overhead. Regarding re-keying method in LKH, OFT, and OKD, the server has more loads for generating, encrypting, and delivering keys to the members. In LKH and OFT, the members do not participate in key calculation on each membership changes. While in OKD the members involve key update process with the key server but the re-keying overhead problem for new members still remains. Conversely in CKCS, the contribution of current members in re-keying process and minimum number of key delivery to new members are two decisive factors which make it more effective than the other ones. Finally, the previously proposed protocols do not consider the simultaneous mode at all. So, the overhead of these approaches is not acceptable for this mode. Assuming simultaneous mode for these protocols, the results are shown in Tables 1, 2, 3 and 4.

In addition, we have implemented programs to compare the overhead of these mentioned protocols. These programs compare the processing time that each protocol spends for generating and encrypting necessary keys after each membership changes based on the number of users.

The source code of the programs is based on the script language program, ruby, using OpenSSL cryptography library. These programs have been run on a 1.66 GHz Windows 7 processor with 2 GB of RAM. We have used AES-256-OFB to generate keys for LKH. The key generation algorithm for OFT, OKD, and CKCSS is based on AES-256-OFB and SHA-1. Finally, we sketch some plots (Figures 11, 12, 13, 14, 15, and 16) for showing numerical comparison. For this purpose, we consider that there are 100,000 group members and 1024, 2048, 4096, and 8192 simultaneous users for join/leave.

### 6.1. Computational Overhead

The computational overhead for these Protocols depend on the number of keys that need to be generated and encrypted by the server. Table 1 shows the key generation overhead at simultaneous join/leave operation. In LKH, group members do not participate in middle node keys calculation in each join/leave operation. In OFT, only a new member at join and the remaining members at leave need to update the keys in their paths. In OKD, when a member joins the group all the necessary keys should be delivered to him/her by unicast and all the remaining members can update their middle node keys by themselves, but at leave some nodes are responsible for updating the affected keys. So,in these protocols, the key generation overhead is $log_2 n$ for a single member and $mlog n$ when $m$ members join/leave the group simultaneously. CKCS has the smallest overhead for key generation comparing with the others.

Table 1.The comparisons of key generation overhead in simultaneous join/leave operations.

| Protocols | Join | Leave |
|-----------|------|-------|
| LKH | $m \log_2 n$ | $m \log_2 n$ |
| OFT | $m \log_2 n$ | $m \log_2 n$ |
| OKD | $m \log_2 n$ | $m \log_2 n$ |
| CKCS | $m+1$ | $1$ |

As mentioned above, the previously proposed protocols do not consider the simultaneous mode. According to the results, in LKH, OFT, and OKD when $m$ members join/leave the multicast group, the server generates keys to update the key tree (all the keys in the paths of $m$ members to the root). If $m=1$, these amounts are equal to the single mode.

In CKCS, this overhead is decreased to $m$ at join and to $1$ at simultaneous leave operation. When m members join the group synchronously, the server generates an individual key for each of them ($m$ individual keys for $m$ members) and one group key. Also, when $m$ members leave the group, the server generates only a new group key for the remaining members. All the necessary keys in CKCS are calculated by the group members.

Table 2 illustrates the key encryption overhead for $m$ simultaneous join/leave. The results show that LKH, OFT and OKD have high overhead at join/leave. But in CKCS, this overhead is the lowest one because in each join operation the server encrypts only the new group key with each new member's individual key. In CKCS, the key encryption overhead at leave is equal to the height of the key tree because the server encrypts the group key by the top node of each part for users who are located at that part.

Table 2.The comparison of key encryption overhead at simultaneous join/leave operations.

| Protocols | Join | Leave |
|-----------|------|-------|
| LKH | $3m \log_2 n$ | $2m \log_2 n$ |
| OFT | $2m \log_2 n$ | $m \log_2 n$ |
| OKD | $m \log_2 n$ | $m \log_2 n$ |
| CKCS | $m$ | $\log_2 n$ |

Figures11 and 12 illustrate the computational overhead (processing time versus the number of simultaneous users) for simultaneous join/leave. As shown, LKH has the highest gradient when *m* members join/leave the group concurrently. OFT and OKD have the lower computational overhead than LKH in simultaneous join/leave. In simultaneous join operation, the computational overhead of OFT is higher than OKD but at leave these protocols have almost the same overhead. CKCSS has the lowest overhead at both simultaneous membership changes.

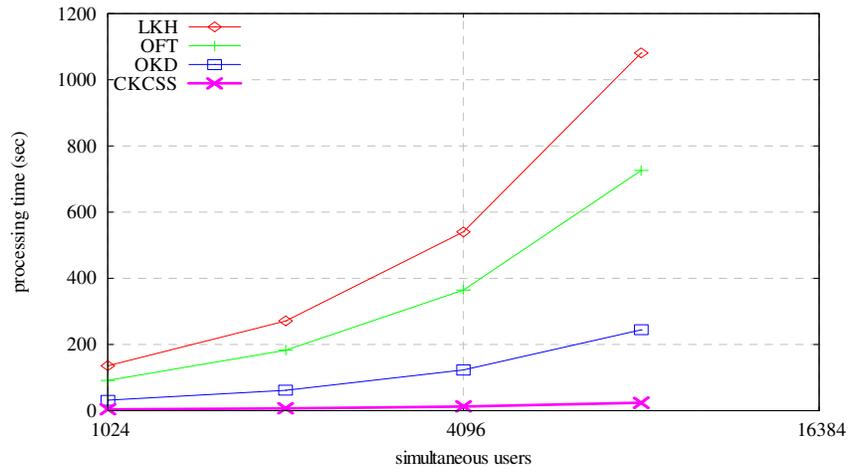

Figure 11. Computational overhead versus number of group members at simultaneous join

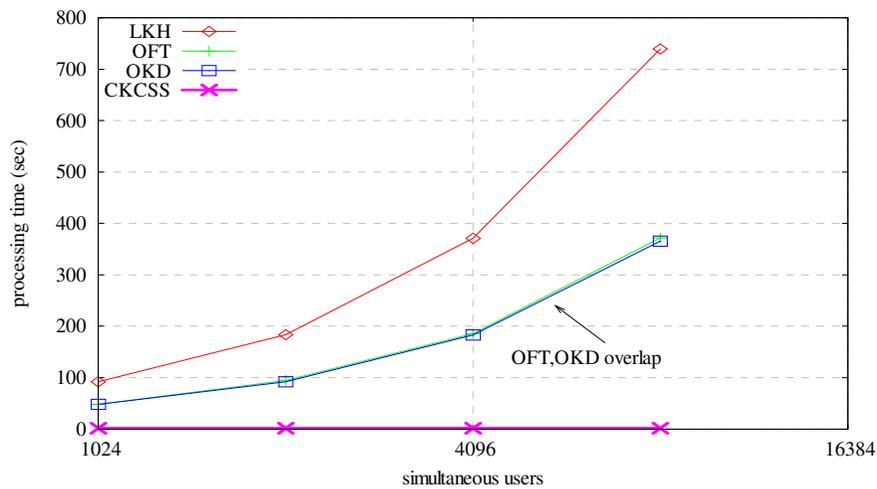

Figure 12. Computational overhead versus number of group members at simultaneous leave

## 6.2. Communication Overhead and Message Size

Table 3 depicts the communication overhead at simultaneous join/leave operation. Communication overhead at simultaneous join is divided into two categories, unicast and multicast overhead. As shown in this table, LKH and OFT have the same communication overhead at join/leave which is the highest one. These two protocols have both unicast and multicast communication in each simultaneous join. This happens because necessary keys for new members are sent by unicast and for remaining members by multicast. In OKD and CKCS, all the keys are collected in one message, and sent to new members by one multicast message. So, there is no overhead for unicast but *1* multicast transmission exists when *m* members join the group.Figures13 and 14 illustrate the numerical results for communication overhead at simultaneous join/leave respectively.

Table 3.The communication overhead at simultaneous join/leave operations.

| Protocols | Join | | Leave |
|---|---|---|---|
| | Unicast | Multicast | Multicast |
| LKH | $n \log_2 n$ | $n\,2\log_2 n$ | $n \log_2 n$ |
| OFT | $n \log_2 n$ | $n \log_2 n$ | $n \log_2 n$ |
| OKD | $n \log_2 n$ | - | $n \log_2 n$ |
| CKCS | - | 1 | $\log_2 n$ |

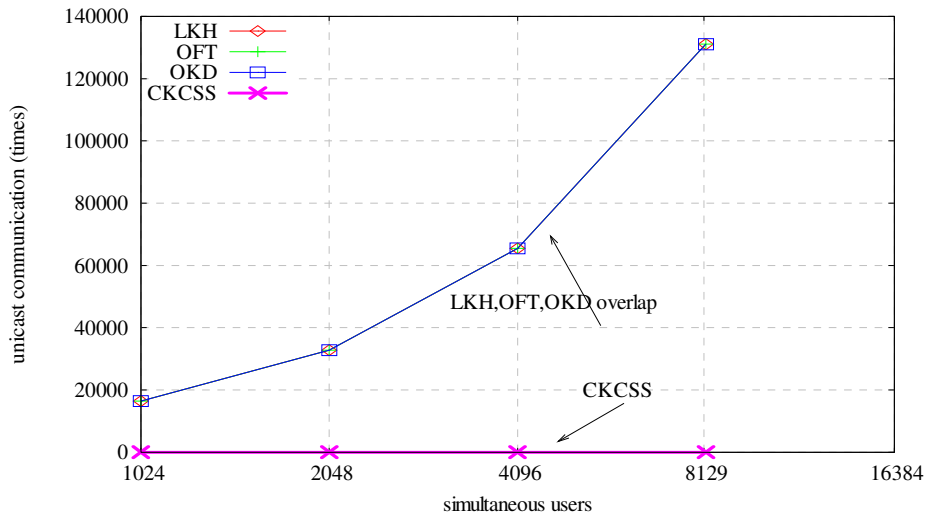

(a)

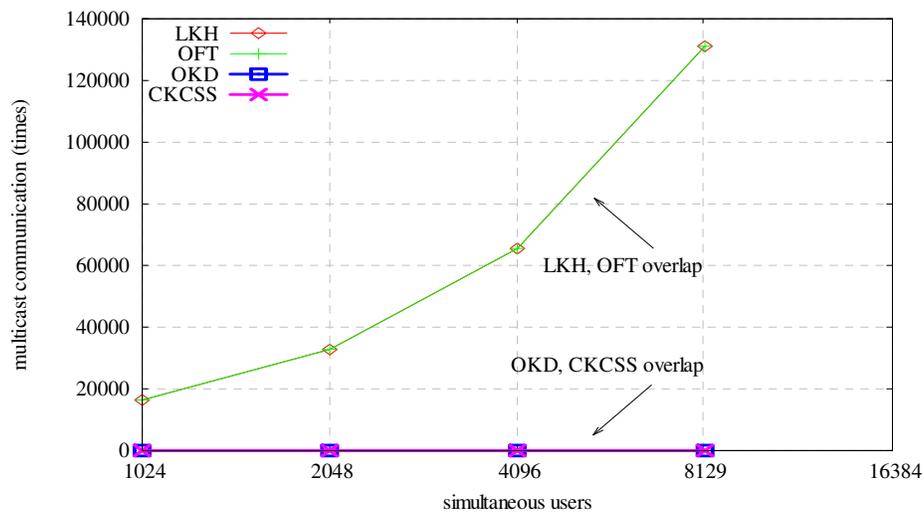

(b)

Figure 13. Communication overhead versus number of group memberssimultaneous join
(a) unicast communication (b) multicast communication

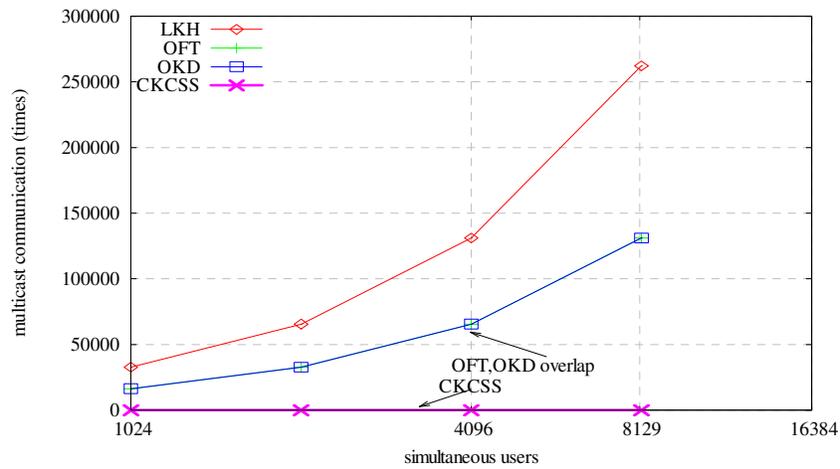

Figure 14. Communication overhead versus number of group members atsimultaneous leave

Table 4 shows message size in simultaneous join/leave. For simultaneous join/leave, CKCS has the lowest message size in each message transmission. All the other protocols have large message size for simultaneous because of their necessary transmissions. In CKCS, the server sends one multicast message which includes $m$ keys. Each of these keys contains the group key encrypted by the individual key of each simultaneous user.Figures15 and 16 illustrate the numerical results for message size at simultaneous join/leave respectively.

Table 4.Message size at simultaneous join/leave operations.

| Protocols | Join | Leave |
|-----------|------|-------|
| LKH | $2m\log_2 n$ | $2m\log_2 n$ |
| OFT | $m\log_2 n +1$ | $m\log_2 n +1$ |
| OKD | $m\log_2 n$ | $m\log_2 n$ |
| CKCS | $m$ | $m$ |

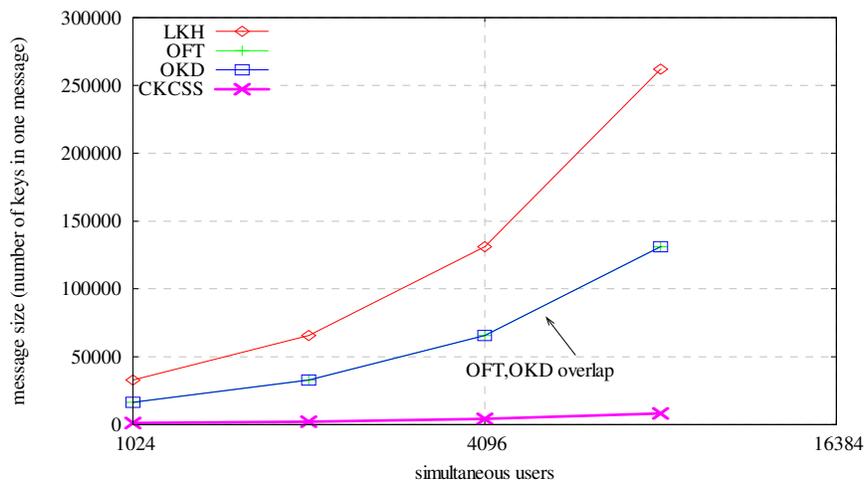

Figure 15.  Message size at simultaneous join

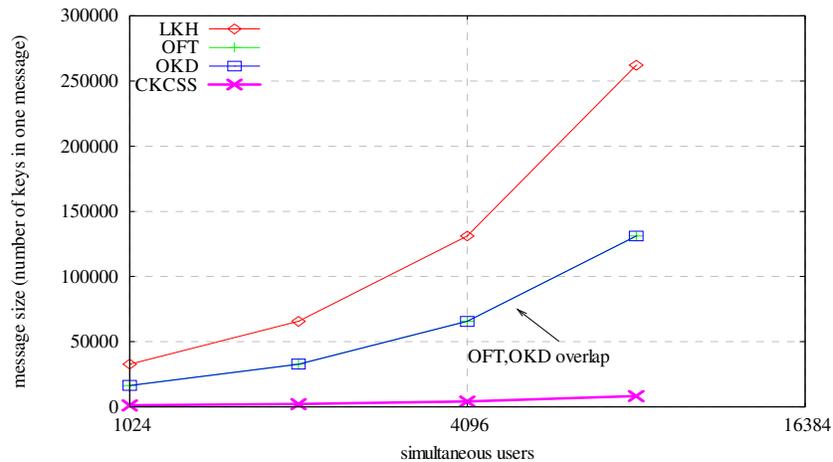

Figure16. Message size at simultaneous leave

Although LKH based protocols minimized the overhead of leave operation to $log_2 n$, they added unnecessary overhead to join operation. This amount gets larger when number of users increases. With a glance at Tables 1, 2, 3, and 4 it is not difficult to see that CKCSS has two major features. First, the overhead of CKCSS at join does not depend on the number of users. It means that the overhead for new member is a constant amount while there is no overhead for current users. Second, reducing the overhead for new user at single join is the other important factor for simultaneous mode. This factor is crucial for simultaneous join.

# 7. CONCLUSIONS

This paper proposed a new group key management protocol, CKCS, for simultaneous join/leave. The protocol is based on logical key hierarchy. In simultaneous mode, when members join the multicast group simultaneously, the server creates a new key tree for the members and their individual keys. The new key tree is attached to the old one by adding a new node to the top of the previous one. When several members leave the group, only the new group key is sent to the remaining members. At the end, we conclude our proposal with some of its contributions:

- CKCS reduces key generation and key encryption overhead largely in simultaneous join/leave.
- CKCS reduces unicast and multicast communication overhead largely at join in simultaneous mode.
- CKCS reduces message size for unicast communication.

**Authors**

**Melisa Hajyvahabzadeh**received her B.Sc. and M.Sc. degree in Information Technology fromSharif University of Technology, International Campus, Iran. Her research intrests include Security, IP Multicast, Group Key Management Protocols, Authentication in various type of networks.

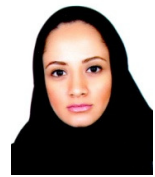

**Elina Eidkhani**received her B.Sc.and M.Sc. degree in Information Technology fromSharif University of Technology, International Campus, Iran. Her research intrests include Security, Wireless Networks, IP Multicast, Group Key Management Protocols in various type of networks.

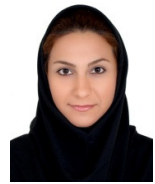

**Seyedeh Anahita Mortazavi**received her B.Sc. and M.Sc. degree in Information Technology fromSharif University of Technology, International Campus, Iran. Her research intrests include Security, IP Multicast, Group Key Management Protocols in various type of networks.

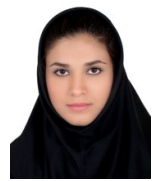

**Alireza Nemaney Pour** has received his B.S degree in Computer Science from Sanno University, Japan, M.S in Computer Science from Japan Advanced Institute of Science And Technology, Japan, and Ph.D. degree in Information Network Science from Graduate School of Information Systems,the University of Electro-Communications, Japan.

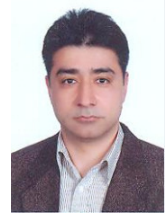

He is currently a faculty memberof Islamic Azad University of Abhar in Iran. In addition, He is a technical advisor of J-Tech Corporation in Japan. His research interests include Network Security, Group Communication Security, Protocol Security, Information Leakage, Spam Mail Prevention, Web Spam Detection, Group Authentication, and Cryptography.